\documentclass[aps,prl,twocolumn,superscriptaddress,msmath,amssymb]{revtex4}

\usepackage{graphicx,bm}
\begin{document}

\title{Neutron scattering study of the oxypnictide superconductor LaO$_{0.87}$F$_{0.13}$FeAs}

\author{Y. Qiu}
\affiliation{NIST Center for Neutron Research, National Institute of Standards 
and Technology, Gaithersburg, MD 20899}
\affiliation{Dept.\ of Materials Science and Engineering, University of
Maryland, College Park, MD 20742}
\author{M. Kofu}
\affiliation{Department of Physics, University of Virginia, Charlottesville, VA 22904}
\author{Wei Bao}
\email{wbao@lanl.gov}
\affiliation{Los Alamos National Laboratory, Los Alamos, NM 87545}
\author{S.-H. Lee}
\affiliation{Department of Physics, University of Virginia, Charlottesville, VA 22904}
\author{Q. Huang}
\author{T. Yildirim}
\author{J. R. D. Copley}
\author{J. W. Lynn}
\affiliation{NIST Center for Neutron Research, National Institute of Standards 
and Technology, Gaithersburg, MD 20899}
\author{T. Wu}
\author{G. Wu}
\author{X. H. Chen}
\affiliation{Hefei National Laboratory for Physical Science at Microscale and Department of Physics, University of Science and Technology of China, Hefei, Anhui 230026, China}

\date{\today}

\begin{abstract}
The newly discovered superconductor LaO$_{0.87}$F$_{0.13}$FeAs ($T_C\approx 26$ K) was investigated using the neutron scattering technique. No spin-density-wave (SDW) order was observed in the normal state nor in the superconducting state, both with and without an applied magnetic field of 9 T, consistent with the proposal that SDW and superconductivity are competing in the laminar materials. While our inelastic measurements offer no constraints on the spin dynamic response from $d$-wave pairing, an upper limit for the magnetic resonance peak predicted from an extended $s$-wave pairing mechanism is provided. Our measurements also support the energy scale of the calculated phonon spectrum which is used in electron-phonon coupling theory, and fails to produce the high observed $T_C$.
\end{abstract}

\pacs{74.70.-b,74.25.Ha,61.05.fg}

\maketitle

A new family of superconductors has been discovered in laminar oxypnictide La(O,F)FeP ($T_C\approx 4$K) \cite{Kamihara2006}, LaONiP ($T_C\approx 3$K) \cite{Watanabe2007}, and La(O,F)FeAs ($T_C\approx 26$K) \cite{Kamihara2008}. Enormous excitement has been generated since $T_C$ was raised above 40 K when La was replaced by Sm ($T_C\approx 43$K) \cite{A033603}, Ce ($T_C\approx 41$K) \cite{A033790},
 Nd ($T_C\approx 52$K) \cite{A034234}, or Pr ($T_C\approx 52$K) \cite{A034283} in $Ln$(O,F)FeAs. Gd(O,F)FeAs is also a superconductor\cite{A034384} and its $T_c$ has been raised to 36 K\cite{A040835}. So far, Sm(O,F)FeAs has the highest $T_C\sim 55$K in the new family of superconductors\cite{A042053,A042105}, and the transition temperature is the highest among all superconductors except in some cuprates.

The parent compounds $Ln$OFeAs ($Ln$=La\cite{Kamihara2008}, Sm\cite{A042105}, Ce\cite{A033790}, Nd, Gd\cite{A034384}) are not superconductors. Instead, a spin-density-wave (SDW) due to Fermi surface nesting develops below $\sim$150 K\cite{A033426,A040795,A040796}. In addition to adding electrons in $Ln$(O,$F$)FeAs,
removing electrons in (La,$Sr$)OFeAs also shifts the Fermi surface out of the nesting condition\cite{A033426,A030429} and leads to superconductors of similarly high $T_C$\cite{wen2008}. Pressure also strongly affects $T_C$\cite{A034266,A041582}, but theoretical calculations do not favor the phonon mechanism\cite{A032740,A031279,A032703}. Various theoretical possibilities involving magnetic fluctuations have been proposed\cite{A032740,A033325,A033982,A034346,A040186,A041739,A041793}. In particular, a pronounced resonance peak in the spin excitations is predicted for an extended $s$-wave superconducting order parameter, while $d$-wave order parameter only modestly enhances spin excitations over the normal state response\cite{A041793}. 

In this neutron scattering study, we chose superconducting La(O,F)FeAs as our subject. Although other $Ln$(O,F)FeAs have higher $T_C$, the La compound is the most thoroughly studied and it also avoids complexity caused by magnetic rare-earth elements. Additionally, the superconducting gap in La(O,F)FeAs has been estimated at $\Delta_0 \approx 3.7(8)$ meV in infrared\cite{A030128}, specific heat\cite{A030928}, and point-contact tunneling\cite{A032405} experiments. The experimental values of $\Delta_0$ set
the theoretical resonance energy of the extended $s$-wave superconductivity at the in-plane antiferromagnetic wavevector ${\rm Q}$=(1/2,1/2,0) and at 5.6$\pm$1.3 meV\cite{A041793}, within the convenient range of neutron scattering spectroscopy. At this moment, only polycrystalline samples of $Ln$(O,$F$)FeAs are available, and an ideal instrument for investigating such samples is the time-of-flight disk chopper spectrometer (DCS) at the NIST Center for Neutron Research. A cryomagnet controlled the sample temperature and magnetic field. The intensity of magnetic inelastic neutron scattering was normalized to absolute units with incoherent nuclear scattering. Our measurements provide valuable insights into the new superconductors and impose experimental constraints for theoretical work.

\begin{figure}[t]
\includegraphics[width=80mm,angle=0,bb=0 0 731 445]{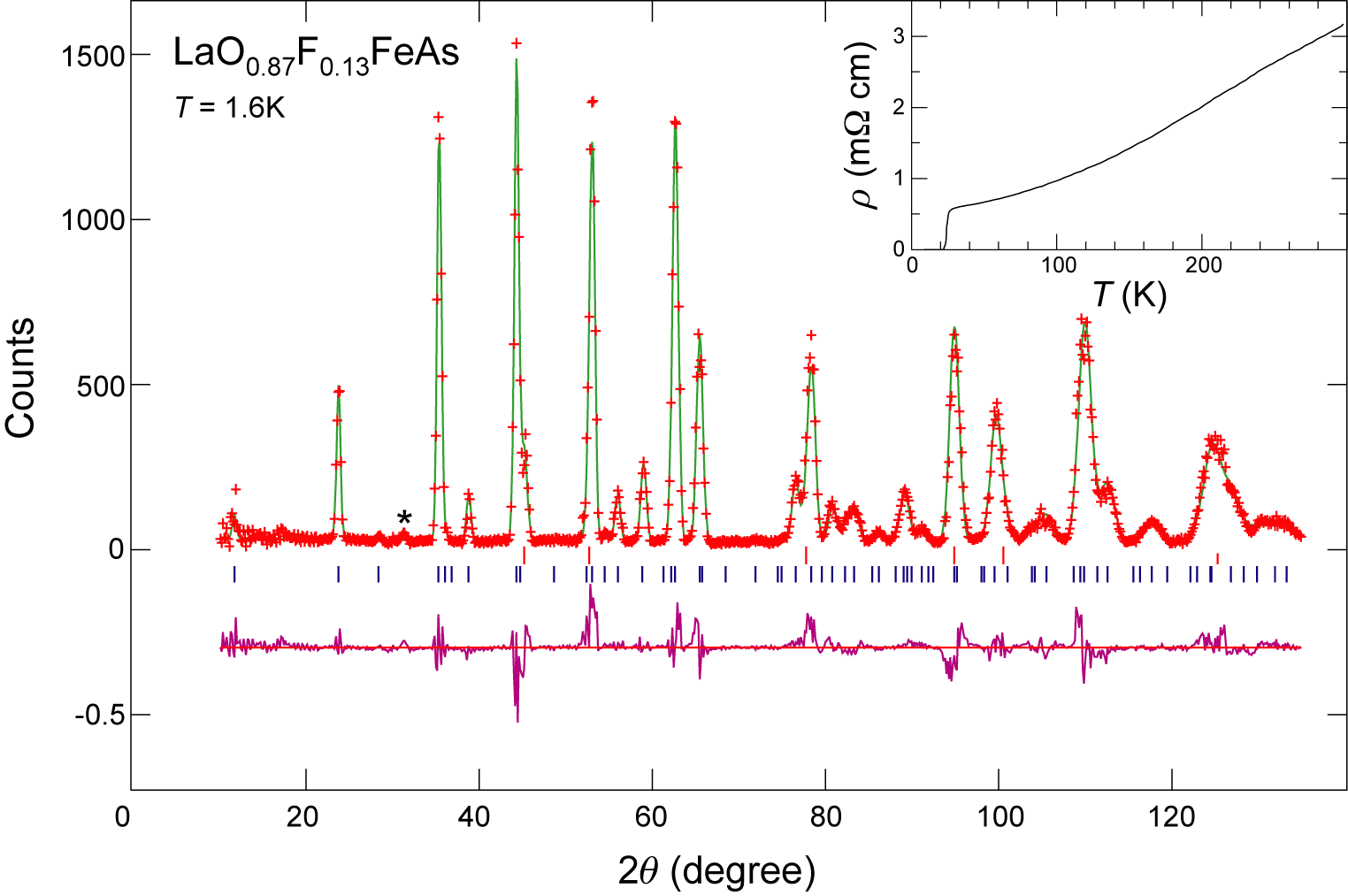}
\vskip -.2 cm
\caption{\label{fig1} (color online)
Observed (crosses) and calculated (solid line) neutron powder diffraction intensities for the superconductor LaO$_{0.87}$F$_{0.13}$FeAs at 1.6 K using space group P4/nmm. Vertical lines are Bragg peak positions for LaO$_{0.87}$F$_{0.13}$FeAs (lower) and Al sample holder (upper).  The data were collected on DCS with an incident beam wavelength $\lambda=1.8\AA$. The structure was refined using the GSAS program\cite{gsas}. Inset: The resistivity of LaO$_{0.87}$F$_{0.13}$FeAs showing the superconducting transition at $T_C=26$ K.}
\end{figure}

A polycrystalline sample of LaO$_{0.87}$F$_{0.13}$FeAs, mass 2 g, was synthesized by the solid state reaction \cite{Kamihara2008}. The observed powder diffraction spectrum at 1.6 K is shown in figure~\ref{fig1}. It is well accounted for by the tetragonal ZrCuSiAs structure as indicated by the refined profile also shown in the figure. 
The sample is of high purity. Only minute impurity peaks are discernible in the diffraction spectrum. Refined structure parameters at 1.6 K are listed in Table~\ref{tab1}.
\begin{table}[b]
\caption{Refined structure parameters of LaO$_{0.87}$F$_{0.13}$FeAs at 1.6 K. 
Space group: P4/nmm (No.\ 129).
 $a=4.0245 (3) \AA$, $c=8.713(1) \AA$, $V=141.126(9) \AA^3$. $Rp$=11.72\%, $wRp$=15.53\%.
}
\label{tab1}
\begin{ruledtabular}
\begin{tabular}{ccllll}
Atom & site & $x$ & $y$ & $z$ & $B(\AA^2)$ \\
\hline
La&2c&1/4&1/4&	0.1442(8)	&	1.0(2)\\
Fe&2b&3/4&	1/4&	1/2&			0.3(1)\\
As&2c&1/4&1/4&		0.6541(8)	&	0.02(2)\\
O/F&2a&3/4&	1/4&	0		&	0.9(2)\\
\end{tabular}
\end{ruledtabular}
\end{table}
Comparing to LaO$_{0.92}$F$_{0.08}$FeAs at a similar temperature\cite{A040795}, both lattice parameters $a$ and $c$ of LaO$_{0.87}$F$_{0.13}$FeAs increase by 0.11\%. Only the As position has a small detectable difference.
The resistivity $\rho$ shown in the inset to Figure~\ref{fig1} was measured using the standard four-probe method.
The superconducting transition starts at $T_C=26$ K, $d\rho/dT$ peaks at 24 K and $\rho$ reaches zero at 22 K. The transition is among the narrowest for
La(O,F)FeAs materials\cite{Kamihara2008,A030128,A030928,A032405,A030623,A031288}. 

\begin{figure}[t]
\includegraphics[width=\columnwidth,angle=0,bb=0 0 397 255]{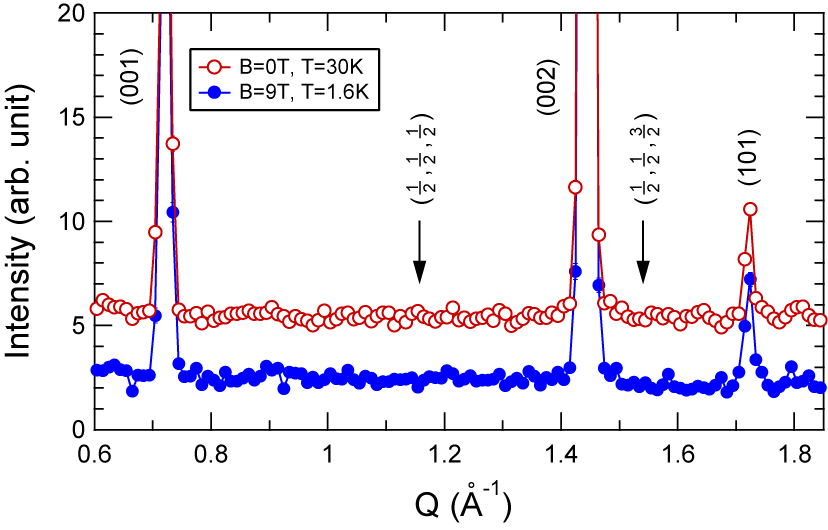}
\vskip -.2 cm
\caption{\label{fig2} (color online)
Neutron powder diffraction intensities of LaO$_{0.87}$F$_{0.13}$FeAs at 30 K and  zero field in the normal state (open red) and at 1.6 K and 9 T magnetic field in the superconducting state (solid blue). The data are collected with a neutron wavelength $\lambda=4.8\AA$ to focus on the small $Q$ range for magnetic signals. The red symbols have been shifted up for clarity. The arrows indicate magnetic Bragg peak positions of the SDW order. No magnetic peak stronger than 0.5\% of the (002) exists in the spectra.}
\end{figure}

The SDW of LaOFeAs is characterized by the wavevector (1/2,1/2,1/2)\cite{A040795,A040796}. The weak staggered magnetic moment $M=0.36(5)\mu_B$ per Fe at 8 K\cite{A040795} can be explained by an associated structure transition\cite{A042252}. The strongest magnetic Bragg peak (1/2,1/2,3/2) is only 1.1\% of the intensity of the structural (002) peak\cite{A040795}. For superconducting samples LaO$_{0.92}$F$_{0.08}$FeAs and LaO$_{0.89}$F$_{0.11}$FeAs, magnetic peaks of the SDW order are not observed above measurement statistics level of about 0.5\% of the (002) peak at 8 K\cite{A040795} and 70 K\cite{A040796}, respectively.
Neither does our superconducting sample LaO$_{0.87}$F$_{0.13}$FeAs show any detectable SDW order down to 1.6 K in the superconducting state (Fig.~\ref{fig1}), nor at 30 K in the normal state (Fig.~\ref{fig2}). Applying a magnetic field of 9 T also does not induce any magnetic peak stronger than 0.5\% of the (002) Bragg peak.
These results support the proposal that the SDW and superconducting order parameters are competing for itinerant electrons and holes on the Fermi surface\cite{A033426,A032740}, and do not favor the theory of coexistence of antiferromagnetism with superconductivity in La(O,F)FeAs. 

Conventional superconductivity is mediated by phonons, and the phonon spectrum has been calculated for La(O,F)FeAs\cite{A030429,A032703}. It has been used to calculate the electron-phonon coupling, and the $T_C$ from this mechanism is much lower than the observed value\cite{A032740,A032703}. To validate the theoretical calculations, we have measured inelastic neutron scattering from phonons in LaO$_{0.87}$F$_{0.13}$FeAs. For polycrystalline samples, the intensity is given by
\begin{equation}
I(Q,\omega)=\sum_{i} \frac{\sigma_i\hbar Q^2}{2m_i} \exp(-2W_i) \frac{{\cal D}_i(\omega)}{\omega}[n(\omega,T)+1],
\label{eq1}
\end{equation}
where $\sigma_i$ and $m_i$ are the neutron scattering cross section and atomic mass of the $i$th atom (La, O/F, Fe, As), $n(\omega,T)$ is the Bose factor, $W_i$ the Debye-Waller factor\cite{royosbn}. The {\em weighted} phonon density of states (PDOS) ${\cal D}(\omega)=\sum_i{\cal D}_i(\omega)$ in Eq.~(\ref{eq1}) differs from the {\em bare} PDOS calculated in \cite{A030429,A032703} by a factor of the squared modulus of phonon eigenvectors. The measured neutron scattering intensity $I(Q,\omega)$ is further weighted by $\sigma_i/m_i$ of different atoms. But the peak positions in the measured and bare PDOS usually remain the same\cite{royosbn}.

\begin{figure}[t]
\includegraphics[width=.85\columnwidth,angle=0,bb=0 0 356 372]{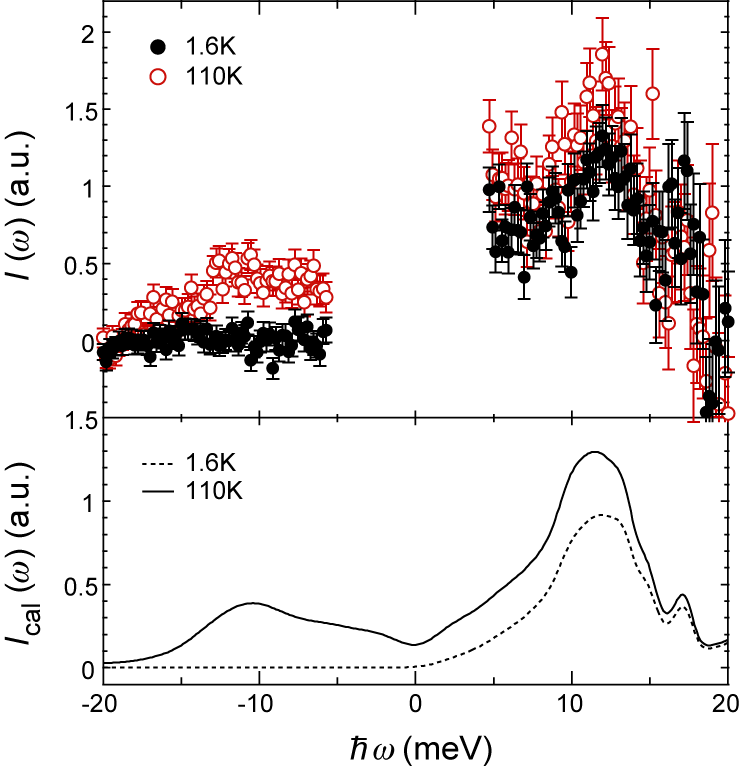}
\vskip 1mm
\includegraphics[width=.85\columnwidth,angle=0,bb=0 0 355 253]{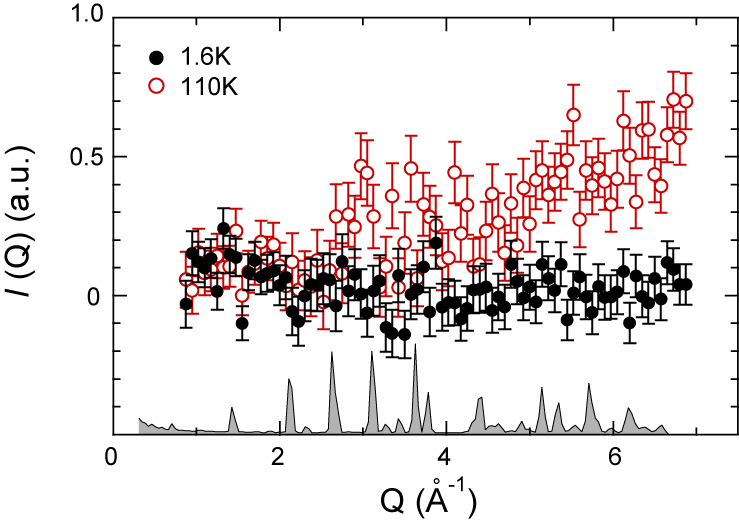}
\caption{\label{fig4} (color online)
Top: $I(\omega)=\int dQ\, S(Q,\omega)$ measured at 1.6 and 110 K. The integration range is from 2.5 to 7 $\AA^{-1}$.
Middle: Calculated {\em bare} intensity profile at 1.6 and 110 K.
Bottom: Measured $I(Q)=\int d\omega\, I(Q,\omega)$ at 1.6 and 110 K. The integration range is from -15 to -5 meV. The shaded profile is measured $S(Q,\omega=0)$.}
\end{figure}
In the top frame of figure~\ref{fig4}, the measured $I(\omega)=\int dQ\, I(Q,\omega)$ at 1.6 and 110 K using neutrons of wavelength 1.8 $\AA$, integrated from $Q$=2.5
to 7$\AA^{-1}$, is shown. In the middle frame is shown the theoretical intensity calculated from the bare PDOS of Singh and Du\cite{A030429}, convoluted with instrument resolution.
At 1.6 K, the Bose factor leads to zero intensity for negative energy transfer, and measurements there serve to determine the background. The integrated intensity $I(Q)=\int d\omega\, I(Q,\omega)$ from the negative energy side, shown in the bottom frame, demonstrates the expected behavior for phonon scattering, which is approximately proportional to $Q^2 I(Q,\omega=0)$\cite{royosbn}. The peak positions of the bare PDOS of Singh and Du are well reproduced in the measured $I(\omega)$. The calculated PDOS in \cite{A032703} closely resembles that in \cite{A030429}. Thus, the phonon spectra used in the
electron-phonon coupling calculations in \cite{A032740,A032703}, which do not favor the phonon mechanism for superconductivity in La(O,F)FeAs, have experimental support from this work.

\begin{figure}[t]
\includegraphics[width=.85\columnwidth,bb=0 0 402 592,angle=0]{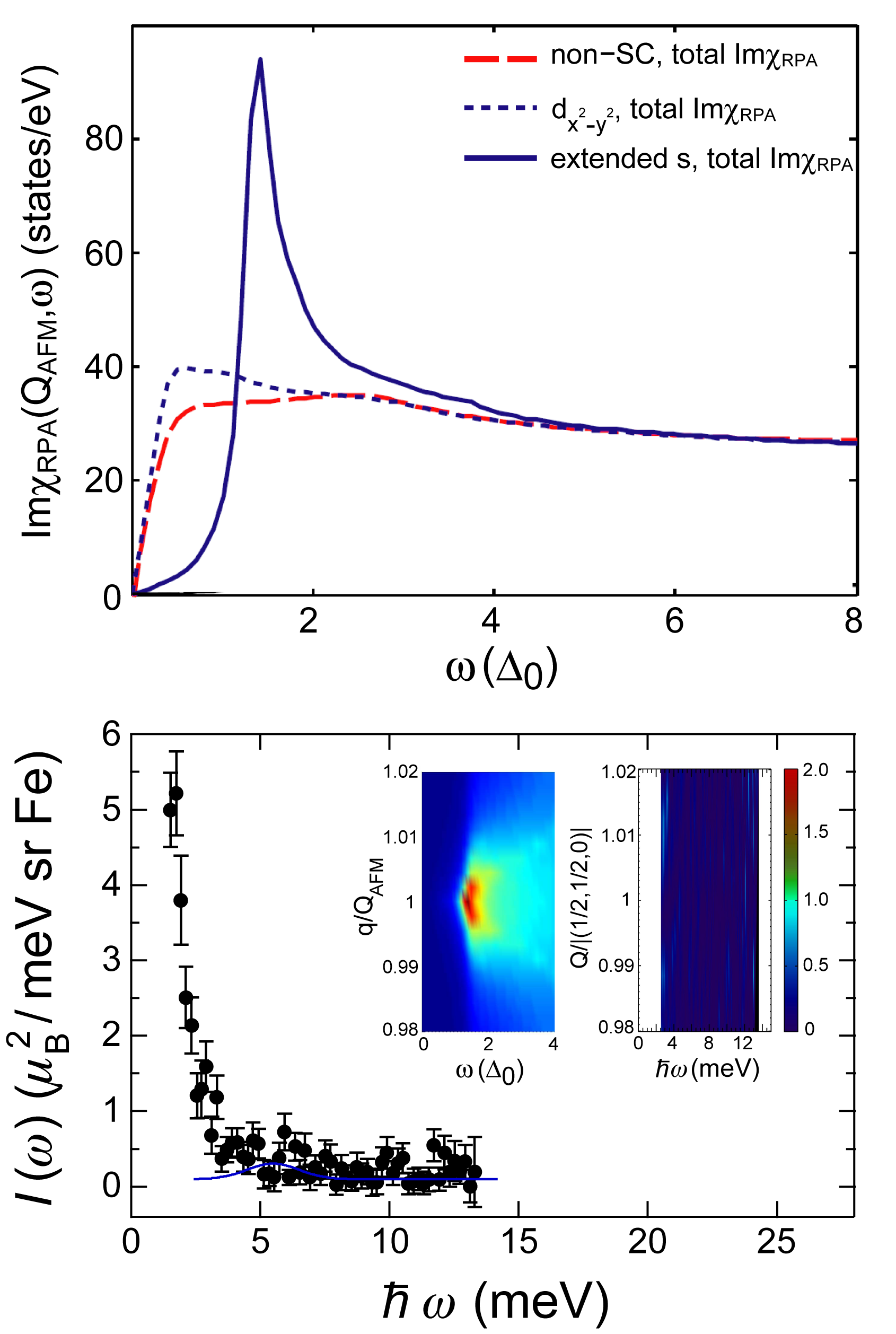}
\vskip -.2 cm
\caption{\label{fig3} (color online)
Theoretical $\chi''({\bm Q},\omega)$ in the superconducting state with an extended $s$-wave order parameter, from \cite{A041793}, is shown in the left inset in the bottom frame. The $\chi''({\bm Q}_{AFM},\omega)$ as a function of $\omega$ in the normal and superconducting state with the two order parameter symmetries\cite{A041793} are shown in the top.
Measured $S({\rm Q},\omega)\equiv \chi''({\rm Q},\omega) \langle n(T,\omega) +1\rangle/\pi \approx \chi''({\rm Q},\omega)/\pi$ at 1.6 K is shown in the right inset in the bottom frame. The color bar indicates intensity in the units of $\mu_B^2$/meV\,sr\,Fe. The $I(\omega)\approx  \chi''({\rm Q}_{AFM},\omega)/\pi$ is shown in the bottom frame.}
\end{figure}

Unconventional superconductivity mediated by various magnetic channels has been investigated theoretically\cite{A032740,A033325,A033982,A034346,A040186,A041739,A041793}. Both $d$-wave and extended $s$-wave have been proposed for the superconducting order parameter of $Ln$(O,F)FeAs. Korshunov and Eremin have investigated the consequences of these pairings in spin dynamics\cite{A041793}. For $d$-wave pairing, the superconducting transition only modestly redistributes the spin spectral weight below $2\Delta_0$, where $\Delta_0$
is the superconducting gap parameter. For extended $s$-wave pairing, a strong resonance peak would appear at the nesting wavevector
${\bm Q}_{AFM}$=(1/2,1/2,0) and $\hbar\omega\sim 1.5\Delta_0$.
In the left inset of the bottom frame and the top frame of Fig.~\ref{fig3}, the theoretical imaginary dynamic spin susceptibility $\chi''({\bm Q},\omega)$ from \cite{A041793} is shown for various cases. 
The value of $\Delta_0$ obtained from infrared measurements is between 3.1 and 3.7 meV\cite{A030128}, in the specific heat study 3.4(5) meV\cite{A030928}, and from tunneling 3.9(7) meV\cite{A032405}. Therefore, the resonance peak at ${\bm Q}_{AFM}$ is between 4.4 and 6.9 meV.

In the right inset to the bottom frame of figure~\ref{fig3}, magnetic neutron scattering intensity $S({\rm Q},\omega)$ measured in the superconducting state at 1.6 K is shown in the same ($\omega, Q$) range as in the left inset. Below 2.5 meV, intensity is dominated by incoherent nuclear neutron scattering and is not shown. The energy dependence at the antiferromagnetic point is shown in the main bottom frame with the same energy scale as in the top frame. Above 2.5 meV, the Bose factor $n(\omega,T)\approx 0$ at 1.6 K. Thus, the measured $S({\rm Q},\omega)\approx \chi''({\rm Q},\omega)/\pi$ can be compared directly to the theoretical $\chi''({\bm Q},\omega)$ in Fig.~\ref{fig3}. Powder averaging will enhance the measured intensity at $Q$ larger than $|{\rm Q}_{AFM}|$ to some extent, however, the sharp resonance peak will be little affected.

We did not observe the strongly enhanced superconducting resonance peak in LaO$_{0.87}$F$_{0.13}$FeAs at 1.6 K. The upper limit for intensity of such a resonance peak is 
0.5(1) $\mu_B^2$/meV\,sr per Fe from our data (see the blue curve in the bottom frame). If the predicted resonance peak is as strong as in the unconventional superconductor CeCoIn$_5$, $\sim$30 $\mu_B^2$/meV\,sr per Co\cite{collin}, being two orders of magnitude stronger than our measurement limit, it would have been observed in our experiments. On the other hand, if the intensity of the resonance peak in La(O,F)FeAs is similar to that in YBa$_2$Cu$_3$O$_{6+x}$, $\sim$0.2 $\mu_B^2$/meV\,sr per Cu\cite{hayden04}, it would not be observed in our measurements. 
Theoretically, the peak intensity for the resonance in La(O,F)FeAs with the extended $s$-wave pairing depends on the choice of damping factor and corrections beyond the random-phase-approximation\cite{A041793}. For the $d$-wave pairing also discussed in \cite{A041793},
the modest change in the spin dynamics would be beyond the sensitivity of this polycrystalline experiment.

In summary, the spin-density-wave order of LaOFeAs is displaced by superconductivity in LaO$_{0.87}$F$_{0.13}$FeAs. The peaks in the theoretical phonon density of states at 12 and 17 meV are observed in our phonon measurements. The theory of phonon mediated superconductivity, which fails to produce the high $T_C\approx 26$ K, thus is based on reliable phonon calculation. Our experiments set an upper limit of 0.5 $\mu_B^2$/meV\,sr per Fe for the resonance peak in the spin excitations at the antiferromagnetic wavevector. Unconventional extended $s$-wave superconductivity mediated by spin fluctuations is constrained by the limit.

We would like to thank D. J. Singh and M.-H. Du for providing theoretical PDOS data. Work at LANL is supported by U.S.\ DOE, at USTC by the Natural Science Foundation of China, Ministry of Science and Technology of China (973 Project No: 2006CB601001) and by National Basic Research Program of China (2006CB922005). The DCS at NIST is partially supported by NSF under Agreement No. DMR-0454672.


\end{document}